\documentclass[preprint]{aastex6}

\def\sm{$\sim$}
\def\e{$e$}
\def\s{$~$}
\def\tm{$\times$}
\def\gcs{g cm$^{-2}$}
\def\GeVN{GeV/n}
\def\TeVN{TeV/n}

\newcommand\tmE[1]{\ensuremath{\times 10^{#1}}}

\usepackage{natbib}

\shorttitle{p and He spectra from CREAM-III}
\shortauthors{Yoon et al.}

\begin{document}

\title{Proton and Helium Spectra from the CREAM-III Flight}

\author{
Y.S.~Yoon\altaffilmark{1,}\footnote{Now at Center for Underground Physics, Institute for Basic Science, Daejeon, 34047, South Korea},
T.~Anderson\altaffilmark{2}, 
A.~Barrau\altaffilmark{3}, 
N.B.~Conklin\altaffilmark{2,}\footnote{Now at Gannon University, 109 University Square, Erie, PA 16541, USA.}, 
S.~Coutu\altaffilmark{2}, 
L.~Derome\altaffilmark{3}, 
J.H.~Han\altaffilmark{1}, 
J.A.~Jeon\altaffilmark{4,}\footnote{Now at Center for Underground Physics, Institute for Basic Science, Daejeon, 34047, South Korea}, 
K.C.~Kim\altaffilmark{1}, 
M.H.~Kim\altaffilmark{1},
H.Y.~Lee\altaffilmark{4},
J.~Lee\altaffilmark{4},
M.H.~Lee\altaffilmark{1,}\footnote{Now at Center for Underground Physics, Institute for Basic Science, Daejeon, 34047, South Korea},
S.E.~Lee\altaffilmark{1},
J.T.~Link \altaffilmark{5,}\footnote{Also at CRESST/USRA, Columbia, MD 21044, USA},
A.~Menchaca-Rocha\altaffilmark{6}, 
J.W.~Mitchell\altaffilmark{5}, 
S.I.~Mognet\altaffilmark{2},
S.~Nutter\altaffilmark{7}, 
I.H.~Park\altaffilmark{4}, 
N.~Picot-Clemente\altaffilmark{1},
A.~Putze\altaffilmark{3}, 
E.S.~Seo\altaffilmark{1,8}, 
J.~Smith\altaffilmark{1},
J.~Wu\altaffilmark{1}}
\altaffiltext{1}{Institute for Physical Science and Technology, University of Maryland, College Park, Maryland, 20742, USA}
\altaffiltext{2}{Department of Physics, Penn State University, University Park, PA 16802, USA}
\altaffiltext{3}{Laboratoire de Physique Subatomique et Cosmologie, Grenoble, France}
\altaffiltext{4}{Department of Physics, Sungkyunkwan University, Suwon 16419, Republic of Korea}
\altaffiltext{5}{Astrophysics Space Division, NASA Goddard Space Flight Center, Greenbelt, MD 20771, USA}
\altaffiltext{6}{Instituto de Fisica, Universidad Nacional Autonoma de Mexico, Mexico}
\altaffiltext{7}{Department of Physics, Northern Kentucky University, Highland Heights, KY 41099, USA}
\altaffiltext{8}{Department of Physics, University of Maryland, College Park, Maryland, 20742, USA}
\begin{abstract}

Primary cosmic-ray elemental spectra have been measured with the balloon-borne Cosmic Ray Energetics And Mass (CREAM) experiment since 2004. 
The third CREAM payload (CREAM-III) flew for 29 days during the 2007-2008 Antarctic season. 
Energies of incident particles above 1 TeV are measured with a calorimeter. 
Individual elements are clearly separated with a charge resolution of \sm0.12 \e\s (in charge units) and \sm0.14 \e\s 
for protons and helium nuclei, respectively, using two layers of silicon charge detectors. 
The measured proton and helium energy spectra at the top of the atmosphere  
are harder than other existing measurements at a few tens of GeV. 
The relative abundance of protons to helium nuclei is 9.53 $\pm$ 0.03 for the range of 1 \TeVN~to 63 \TeVN. 
The ratio is considerably smaller than other measurements at a few tens of GeV/n. 
The spectra become softer above \sm 20 TeV. 
However, our statistical uncertainties are large at these energies and more data are needed.
\end{abstract}

\keywords{}

\section{Introduction} \label{sec:intro}

A hardening of TeV elemental energy spectra was first reported \citep{2010ApJ...714L..89A} using results from the CREAM-I flight. 
The more abundant than expected fluxes of protons and helium nuclei in the TeV region  were verified by later experiments,
such as the Payload for Antimatter Matter Exploration and Light-nuclei Astrophysics \citep[PAMELA]{2013AdSpR..51..219A} and 
the Alpha Magnetic Spectrometer-02 \citep[AMS-02]{2015PhRvL.115u1101A,2015PhRvL.114q1103A}. 
The hardening and higher abundances are difficult to explain using general models of cosmic-ray acceleration 
\citep{1977DoSSR.234.1306K,1978MNRAS.182..147B,1978ApJ...221L..29B}, 
and their transport in the Galaxy, which predict single power-law cosmic-ray spectra. 
Possible explanations for the hardening and higher abundance have been suggested, 
including but not limited to,
effects of the cosmic-ray source spectrum \citep{2010ApJ...725..184B,2013ApJ...763...47P}, 
effects due to propagation through the Galaxy \citep{2012PhRvL.109f1101B,2013JCAP...07..001A}, 
or the effect of nearby local sources \citep{2012MNRAS.421.1209T,2012APh....35..449E,2013A&A...555A..48B}.
Other possible causes include cosmic-ray re-acceleration by weak shocks in the Galaxy \citep{Ptuskin2011icrc, 2014A&A...567A..33T},
injection processes in collisionless shock acceleration \citep{2012PhRvL.108h1104M,1997ApJ...487..197E},
and inhomogeneous ambient medium sources, called superbubbles \citep{2011MNRAS.415.1807D, 2011ApJ...729L..13O,2016PhRvD..93h3001O}.
Furthermore, CREAM-I results were utilized by a recent model interpreting air-shower measurements 
above \sm1000 TeV \citep{2013FrPhy...8..748G}.
More accurate measurements above a few TeV would be helpful not only to understand elemental energy spectra in the TeV energy region, 
but also to provide parameters for fitting data from ground measurements.

The balloon-borne CREAM experiment has measured elemental spectra of cosmic rays from protons to iron nuclei since 2004. 
It has flown six times circumnavigating the Antarctic continent with \sm161 days of accumulated exposure 
\citep{2014AdSpR..53.1451S}. Proton and helium energy spectra from the third flight (CREAM-III) are discussed here.

\section{CREAM-III Experiment} \label{sec:exp}
\subsection{CREAM Flight 2007-2008} \label{sec:flight}
The CREAM-III payload had a successful 29-day flight around the Antarctic continent. 
It was launched on a long-duration balloon from Williams Field near McMurdo station, Antarctica on December 19, 2007,
and it landed on January 17, 2008. The float altitude during the flight was between 37 and 40 km with an average atmospheric overburden of 3.9 $\pm$ 0.4 \gcs. 
The instrument was controlled from a science operation center (SOC) at the University of Maryland throughout the flight. 
Primary uplink was via the Tracking and Data Relay Satellite System (TDRSS) with a backup connection via Iridium satellites. 
All the prime data including science events and housekeeping events were transmitted in near real-time through TDRSS via a high-gain antenna (up to \sm85 kbps). 
All of the data were also stored in an on-board disk on the science flight computer. 
About 30 GB of data were collected, including data transmitted to the SOC and data stored on the on-board disk.
 
The flight operation went smoothly and the detectors performed well. 
For instance, the calorimeter did not need any further in-flight adjustments 
for trigger thresholds or high-voltage settings of its hybrid photodiodes, 
whereas those were adjusted in the previous two flights to optimize data rates. 
The live-time fraction during data collection was about 99\%. 
About $1.3 \times 10^{6}$ science events were collected during the flight.
Significantly more events were collected around 1 TeV, 
because a reduced noise level in the calorimeter electronics allowed the trigger thresholds to be lowered to about 15 MeV. 
In addition, a sparsification threshold value in each calorimeter channel, which suppresses the pedestal, 
was lowered to about 2 $\sigma$ of the pedestal value itself.

\section{CREAM Instrument} \label{sec:inst}
\begin{figure}
\centering
\includegraphics[width=0.8\textwidth]{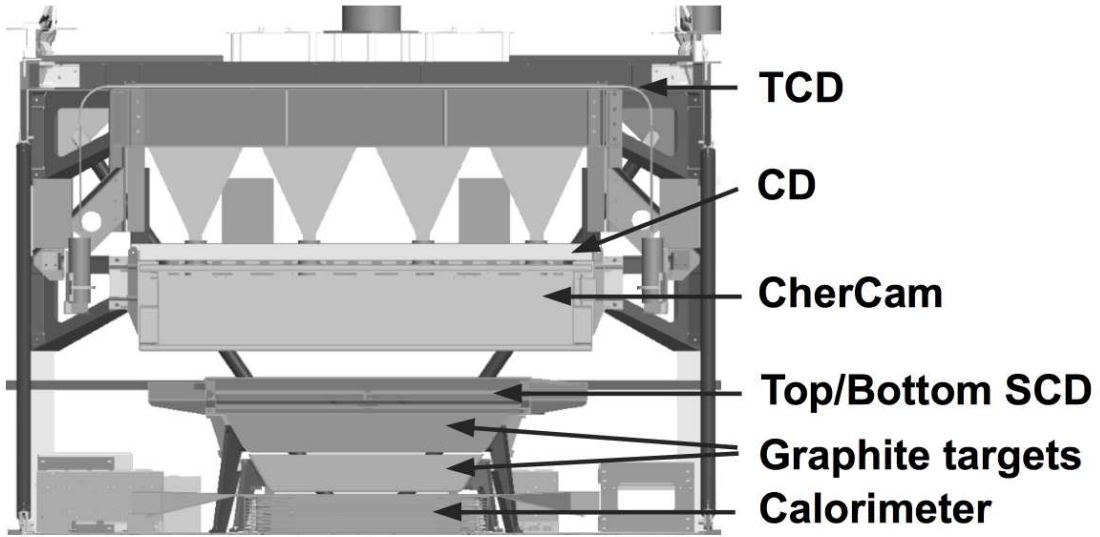}
\caption{CREAM-III instrument; from the top: TCD, CD, CherCam, Top/Bottom SCD, graphite target blocks, and Calorimeter.}
\label{fig:instrument}
\end{figure}

The CREAM-III instrument included a tungsten/scintillating fiber calorimeter, 
a dual layer Silicon Charge Detector (SCD), a Cherenkov Camera (CherCam), 
a Cherenkov Detector (CD), and a Timing Charge Detector (TCD), as shown in Fig. \ref{fig:instrument}. 
The calorimeter measures the energy of incident nuclei that interact in graphite targets located directly above it. 
The calorimeter is comprised of a stack of 20 tungsten layers with an overall 20-radiation length (X$_{0}$) depth and 20 layers of scintillating fibers. 
The calorimeter for the CREAM-III flight had the same detector configuration as the calorimeter in the previous two flights 
with reduced electronics noise levels and finer digitization than for the previous two flights \citep{2009ITNS...56.1396L}. 
The dual layer SCD, CherCam, and TCD provided redundant charge identification of incident particles by measuring dE/dx in the silicon layers, 
Cherenkov rings in the CherCam, and dE/dx in the plastic scintillators, respectively. 
The dual layer SCD provided two independent charge measurements, the same as in the second flight \citep{2007ITNS...54.1743N}. 
The CherCam, inaugurated with the CREAM-III flight, and the TCD provided redundant charge measurements 
by measuring the Cherenkov signal from the aerogel radiator \citep{2011JInst...6.6004B} and 
measuring dE/dx in plastic scintillators \citep{2007NIMPA.572..485C}, respectively.  
Besides sub-detectors, the Science Flight Computer (SFC) was upgraded employing a redundant architecture 
based on a high speed USB 2.0 bus \citep{2008ICRC....2..401Y}. 
Its stability and robustness had been verified through lab test and flights \citep{2009ICRCHanSK}. 
In this analysis of proton and helium energy spectra, an energy measurement 
from the calorimeter and two independent charge measurements from the top and bottom SCD are 
used because the SCD provides the best charge measurements with a resolution of \sm0.12 e and \sm0.14 for protons and helium nuclei, respectively. 
In addition, the geometric acceptance between the calorimeter and the SCD is 
larger than that between the calorimeter and other sub-detectors for charge measurements.   

\section{Data Analysis} \label{sec:analy}
The procedures followed in the proton and helium spectra analysis with CREAM-III data 
are consistent with the procedures used in analyzing the first flight data \citep{2011ApJ...728..122Y}: 
event selection, charge determination, energy measurements, spectral deconvolution, 
background corrections, absolute fluxes, and uncertainties were handled in a manner comparable 
to previous work. Updated methods and some differences are described below.   

The CREAM-III instrument had two main trigger systems for science events:  
significant energy deposits in the calorimeter for high-energy particles or large pulse height, Z$>$2, in the TCD for heavy nuclei. 
They were similar to those used in previous flights, but the trigger threshold for the calorimeter was set lower than in previous flights. 
To be explicit, the calorimeter trigger system required each of any six consecutive layers in the calorimeter 
to have at least one ribbon with a deposit of more than 15 MeV, 
whereas a value of 45 MeV had been used in the first flight, resulting in measurements around 1 TeV. 
The high-energy shower events that meet the calorimeter trigger condition were used in this analysis.

\subsection{Event Selection} \label{subsec:evtsel}
The shower axis was reconstructed with a least-squares fit to a straight line in the XZ and YZ planes \citep{2007NIMPA.579.1034A,2011ApJ...728..122Y}. 
The resulting trajectory resolution (1$\sigma$) was  \sm0.9 cm on the top SCD plane. 
The reconstructed trajectory required traversing the top and bottom SCD active areas and the bottom of the calorimeter area. 
Events with late interactions, where an electromagnetic shower was generated 
by the primary cosmic ray starting in the middle or bottom layers of the calorimeter 
were removed because they could result in an underestimation of deposited energy or charge misidentification 
due to uncertainties in trajectory reconstruction. 

\subsection{Charge Determination} \label{subsec:zdeter}
\begin{figure}
\centering
\includegraphics[width=0.8\textwidth]{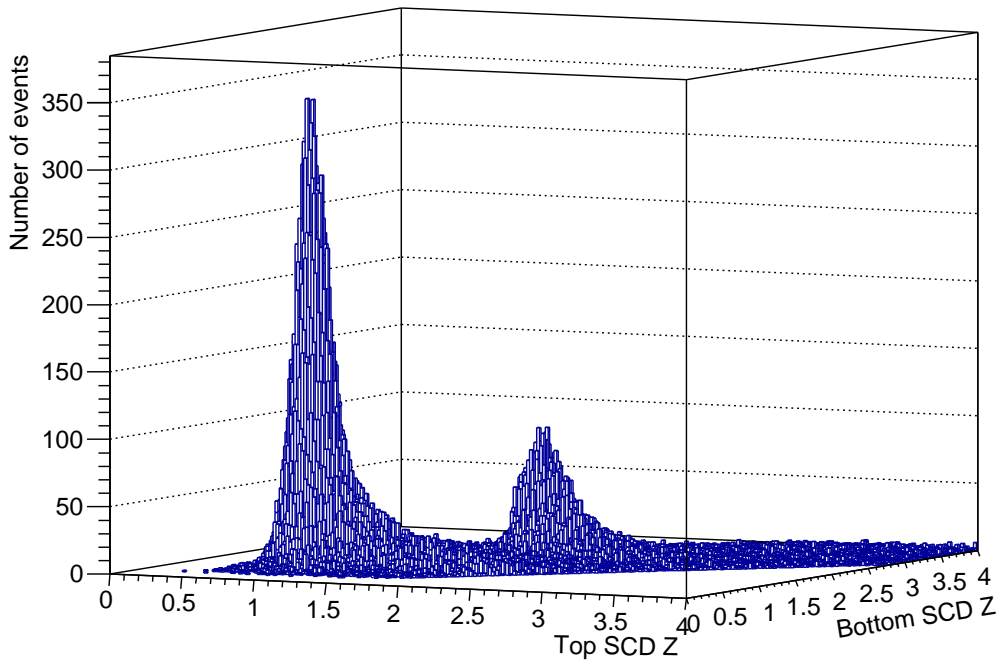}
\caption{Charge (Z) distribution from the top and bottom SCD. Consistency between the top and bottom SCD charge was required.}
\label{fig:tbscdz}
\end{figure}

Two independent charge measurements from the top and bottom SCD layers were used for charge identification,
which is a significant update in the analysis compared to the procedures followed for CREAM-I data analysis. 
Charge identification of an incident particle was determined by scanning a 7\tm7 pixel area, 
centered on the extrapolated hit position from the reconstructed calorimeter track, 
to seek the highest pixel signal in each top and bottom SCD layer. 
The signal from the SCD pixel reflects the ionization energy loss per unit path length (dE/dx) of an incident particle 
in the thin silicon layer (380 $\mu$m), the energy loss being proportional to  Z$^{2}$. 
The expected contamination by back-scattered particles from the calorimeter was less than 3\% 
when charge identification with reconstructed trajectory was used \citep{2008ICRC....2..381P}. 
Two independent charge measurements and reduced noise level in the top and bottom SCD electronics 
allowed for a clear separation between proton and helium peaks, as shown in Fig. \ref{fig:tbscdz}. 
The charge resolution is estimated to be \sm0.12 $e$ for protons and \sm0.14 $e$ for helium nuclei, respectively. 
Events with 0.5$<$Z$_{\rm{topSCD}}<$1.55 and 0.5$<$Z$_{\rm{bottomSCD}}<$1.55 were selected 
as protons and events with 1.65$<$Z$_{\rm{topSCD}}<$2.65 and 1.65$<$Z$_{\rm{bottomSCD}}<$2.65 were selected as helium nuclei. 
The loss of proton and helium events located far in the dE/dx Landau tails due to the applied charge-range selection 
was corrected by applying a charge-selection efficiency, which will be discussed in Section \ref{subsec:bkgs}.  

\subsection{Energy Measurement} \label{subsec:emea}
The calorimeter was calibrated with a 150 GeV electron beam at CERN before flight \citep{2009ICRCHan}. 
The calorimeter calibration includes an accounting of the ratio of MeV to Analog-to-Digital Conversion (ADC) units, 
a gain correction due to high-voltage setting differences, and a range-correction value. 
A ratio of MeV to ADC for each ribbon was obtained using signals from beam test data and Monte Carlo (MC) simulations results, 
generated by GEANT/FLUKA 3.21 \citep{Brun1984,Fasso1993}. 
Measured signals in ADC units were corrected for gain differences due to different high-voltage settings 
on the calorimeter's hybrid photodiodes between the beam test and flight. 
More details can be found in \citet{2008ICRC....2..421Y,2006NuPhS.150..272A,2009ICRCHan}.

\subsection{Spectral Deconvolution} \label{subsec:decon}
The distribution measured by the calorimeter was deconvolved into an incident energy distribution using matrix relations \citep{2011ApJ...728..122Y}. 
The counts, N$_{inc,i}$ in the i$^{th}$incident energy bin were estimated from the measured counts, N$_{dep,j}$ in deposited energy bin j by the relation

\begin{equation}
N_{inc,i}= \sum_{j} P_{ij} N_{dep,j}, 
\end{equation}
where matrix element P$_{ij}$ is the probability that the events in the deposited energy bin j are from the incident energy bin i. 
The matrix elements P$_{ij}$ were estimated from the response matrix generated by MC simulation results obtained separately 
for protons and helium nuclei with CREAM-III flight settings. 
The spectral deconvolution method for CREAM-III analysis is the same as that used in the CREAM-I analysis. 
Only response matrices from MC simulations were updated. More details about MC simulations are in \citet{2011ApJ...728..122Y}.    

\subsection{Backgrounds} \label{subsec:bkgs}
The primary background is due to events with misidentified charge, which resulted from secondary particles generated by interactions above the SCD or particles backscattered from the targets and calorimeter. 
Numbers for misidentified proton and helium events were estimated from MC simulations with a power-law  spectrum. 
About 5.5\% of measured protons were actually misidentified helium and about 5.8\% of measured helium nuclei were misidentified protons. 
Less than 1\% of triggered protons and helium nuclei were from secondary particles. 
Another source of background comes from out-of-geometry events, which are not passing through the SCD active area, 
but have a reconstructed trajectory within the SCD active area.
According to MC simulations, the background due to reconstruction is about 1.2\% and 2.5\% for protons and helium nuclei, respectively.

\subsection{Absolute Flux} \label{subsec:flux}
The measured spectra are corrected for the instrument acceptance and efficiencies to obtain the absolute flux F:
\begin{equation}
F=\frac{dN} {dE} \cdot \frac{1}{GF ~\epsilon ~T ~\eta},
\end{equation}
where dN is the number of events in an energy bin, dE is the energy bin size, GF is the geometry acceptance factor, 
$\epsilon$ comprises the efficiencies, T is the live time, and $\eta$ is the survival fraction accounting for atmospheric attenuation.
The geometric acceptance was estimated to be 0.322 m$^2$ sr 
by requiring the extrapolated calorimeter trajectory to traverse the top and bottom SCD active areas and the bottom of the calorimeter. 
During data collection, the live-time fraction was \sm99\%, and data were collected for 1.80\tmE{6} s, 
which represents about 21 of the 29 days of flight, excluding initial instrument tuning processes and routine pedestal and calibration processes. 
Efficiencies were estimated with MC simulation results for protons and helium nuclei. 
The trigger efficiency, i.e., the fraction of events satisfying the trigger condition 
among all events passing both SCDs and calorimeter active areas, 
was 76\% for protons and 89\% for helium nuclei, respectively. 
The reconstructed efficiency, i.e., the fraction of events satisfying the reconstruction condition 
among triggered events, was 97.5\% for protons and 98.4\% for helium nuclei, respectively. 
The efficiency factor resulting from removing events with a late interaction in the calorimeter 
was 83.8\% for protons and 93.8\% for helium nuclei, respectively. 
The charge efficiency, i.e., the ratio of events with successfully reconstructed charge 
among all reconstructed events, excluding events with late interaction, 
was 60.9\% for protons and 58.3\% for helium nuclei, respectively. 
This charge efficiency includes a 78\% efficiency factor due to masking unstable top and bottom SCD channels.
The estimated survival fraction was corrected for each event to account for the atmosphere above the instrument and 
the instrument material depth along the reconstructed trajectory. 
The average survival fraction for the average atmospheric depth of 3.9 \gcs, 
and the instrument material thickness above the SCDs of 12 \gcs, at the average incidence angle of 35 degrees, 
is 81\% and 69\% for protons and helium nuclei, respectively. 

\subsection{Uncertainties} \label{subsec:uncer}
The statistical uncertainty in each energy bin was estimated given its number of entries, 
by considering the 68.3\%  Poisson confidence interval determined by \citet{1998PhRvD..57.3873F}. 
Systematic uncertainties associated with the efficiencies and background 
were estimated based on MC simulations, in order to account for energy-dependent effects. 
The uncertainty due to the trigger efficiency was about 1.9\% for protons and 2.4\% for helium nuclei, respectively. 
The uncertainty associated with the reconstruction efficiency, removing late shower events, 
and the charge selection efficiency amounted to less than 1\%. 
The uncertainty on the background from event reconstruction was 1.8\% and 2.6 \% for protons and helium nuclei, respectively. 
The uncertainty on the background due to misidentified charge was 1\% for protons and helium nuclei, respectively. 
The systematic uncertainty on the survival fractions in the atmosphere was calculated analytically, 
as discussed in \citet{2011ApJ...728..122Y}, and amounted to 2\% and 3\% for protons and helium nuclei, respectively. 
Finally, the uncertainty associated with the energy measurement was about 10\% due to energy calibration.
  
\floattable
\begin{deluxetable}{ccc}
\tablecaption{CREAM-III proton fluxes \label{tab:creamiiip}}
\tablecolumns{3}
\tablenum{1}
\tablewidth{0pt}
\tablehead{
\colhead{$E_{i}$ } &
\colhead{$E_{f}$ } &
\colhead{Flux}\\
\colhead{(GeV) } & 
\colhead{ (GeV) } &
\colhead{(m$^{2}$ s sr GeV)$^{-1}$}
}
\startdata
$10^{3}$    & 1.58\tmE{3} & $ ( 4.07 \pm 0.05) \times 10^{-5} $ \\
1.58\tmE{3} & 2.51\tmE{3} & $ ( 1.21 \pm 0.02 ) \times 10^{-5} $ \\
2.51\tmE{3} & 3.98\tmE{3} & $ ( 3.72 \pm 0.10 ) \times 10^{-6} $ \\
3.98\tmE{3} & 6.31\tmE{3} & $ ( 1.16 \pm 0.05 ) \times 10^{-6} $ \\
6.31\tmE{3} & $10^{4}$    & $ ( 3.63 \pm 0.02 ) \times 10^{-7} $ \\
$10^{4}$    & 1.58\tmE{4} & $ ( 1.10 \pm 0.09 ) \times 10^{-7} $ \\
1.58\tmE{4} & 2.51\tmE{4} & $ ( 3.14 \pm 0.38  ) \times 10^{-8} $ \\
2.51\tmE{4} & 3.98\tmE{4} & $ ( 8.21 \pm 1.50   ) \times 10^{-9} $\\
3.98\tmE{4} & 6.31\tmE{4} & $ ( 1.99 \pm 0.60  ) \times 10^{-9} $\\
6.31\tmE{4} & $10^{5}$    & $ 5.9^{+3.2}_{-1.7} \times 10^{-10} $\\
$10^{5}$    & 1.58\tmE{6} & $ 1.6^{+1.6}_{-1.0} \times 10^{-10} $ \\
1.58\tmE{6} & 2.51\tmE{6} & $ 4.5^{+2.8}_{-1.3} \times 10^{-11} $ \\
\enddata
\end{deluxetable}

\floattable
\begin{deluxetable}{ccc}
\tablecaption{CREAM-III helium fluxes \label{tab:creamiiihe}}
\tablecolumns{3}
\tablenum{2}
\tablewidth{0pt}
\tablehead{
\colhead{$E_{i}$} & 
\colhead{$E_{f}$ } &
\colhead{Flux}\\
\colhead{(\GeVN) } &
\colhead{ (\GeVN) } &
\colhead{(m$^{2}$ s sr GeV n$^{-1}$)$^{-1}$}
}
\startdata
2.51\tmE{2} & 3.98\tmE{2}     & $ (1.38 \pm 0.02 ) \times 10^{-4} $ \\
3.98\tmE{2} & 6.31\tmE{2}     & $ (4.35 \pm 0.08 ) \times 10^{-5} $ \\
6.31\tmE{2} & $10^{3}$ 	     & $ (1.35 \pm 0.03 ) \times 10^{-5} $ \\
$10^{3}$      & 1.58\tmE{3}     & $ (4.26 \pm 0.15 ) \times 10^{-6} $ \\
1.58\tmE{3} & 2.51\tmE{3}     & $ (1.30 \pm 0.07 ) \times 10^{-6} $ \\
2.51\tmE{3} & 3.98\tmE{3}     & $ (3.83 \pm 0.29 ) \times 10^{-7} $ \\
3.98\tmE{3} & 6.31\tmE{3}     & $ (1.16 \pm 0.13 ) \times 10^{-7} $ \\
6.31\tmE{3} & $10^{4}$ 	     & $ (3.25 \pm 0.54 ) \times 10^{-8} $ \\
$10^{4}$      & 1.58\tmE{4}     & $ (7.95 \pm 2.10 ) \times 10^{-9} $ \\
1.58\tmE{4} & 2.51\tmE{4}     & $  2.4^{+0.9}_{-0.8}\times 10^{-9} $ \\
2.51\tmE{4} & 3.98\tmE{4}     & $  8.1^{+2.9}_{-2.1} \times 10^{-10} $\\
3.98\tmE{4} & 6.31\tmE{4}     & $  1.7^{+2.4}_{-1.0} \times 10^{-10} $\\
\enddata
\end{deluxetable}

\section{Results}
\begin{figure}
\centering
 \includegraphics[width=0.48\textwidth, trim= 10 10 110 10]{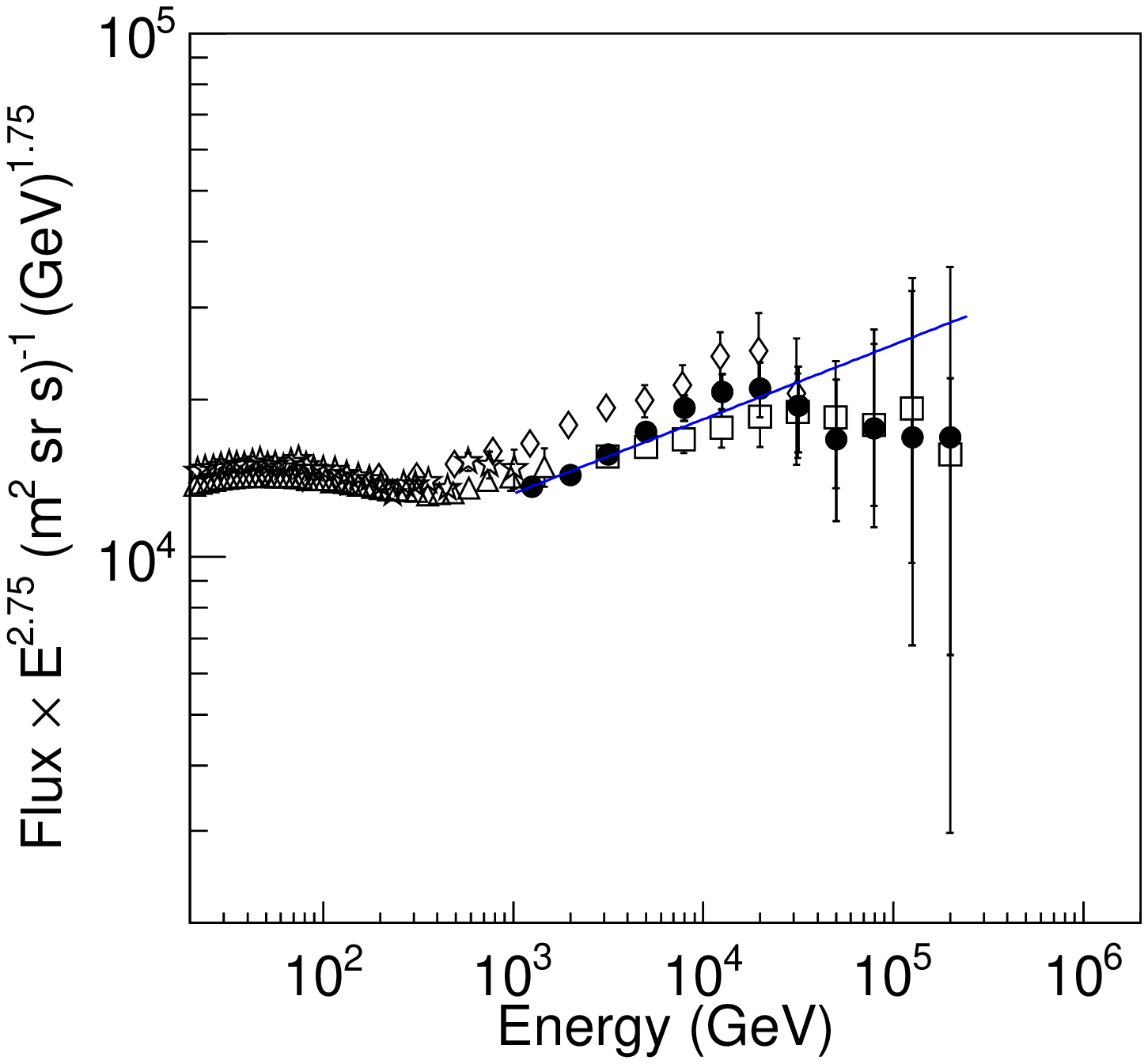}
 \includegraphics[width=0.48\textwidth, trim= 10 10 110 10]{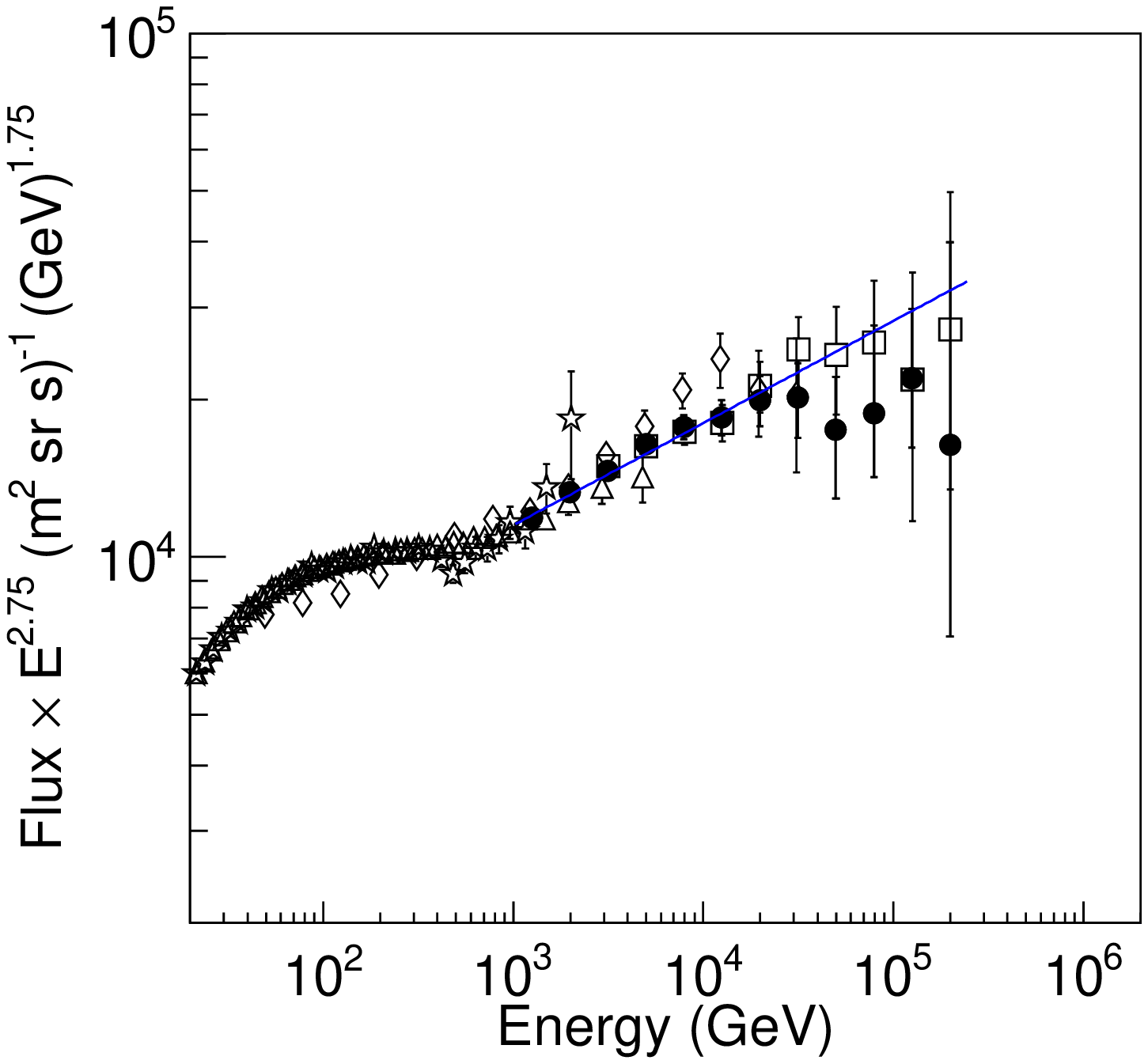}\\
\caption
{Proton (left) and helium (right) spectra from CREAM-III (filled circles) with power-law fits (lines). Statistical uncertainties are shown. 
Selected previous measurements are also shown: AMS-02 (triangles), ATIC-2 (diamonds), CREAM-I (squares), and PAMELA (stars).}
\label{fig:hheflux}
\end{figure}

The measured proton and helium fluxes in the energy range of 1 TeV to 250 TeV
are given in Tables \ref{tab:creamiiip} and \ref{tab:creamiiihe}, respectively, are shown in Fig. \ref{fig:hheflux} and
are also compared to selected previous measurements: 
CREAM-I, Advanced Thin Ionization Calorimeter \citep[ATIC]{2009BRASP..73..564P}, 
PAMELA, and AMS-02. 
The CREAM-III proton and helium spectra are in agreement with other measurements near 1 TeV, 
such as PAMELA and AMS-02, as shown in Fig. \ref{fig:hheflux}. 
The proton flux of CREAM-III near ~1.3 TeV shows an overlap with AMS-02. 
The helium flux of CREAM-III shows an overlap with AMS-02 and PAMELA results in the region of their maximum energy reach. 
The measured proton and helium spectra of CREAM-III are in agreement with CREAM-I results within uncertainties, 
over the energy range of 2.5 TeV to 250 TeV.

Power-law fits for the proton and helium spectra are represented by 
\begin{equation}
\frac{d\Phi} {dE} = \Phi_{0}~E^{-\alpha}~ \rm{ (m^{2}~sr~s~GeV~n^{-1})^{-1}}.
\end{equation}

The fit parameters for protons and helium nuclei from 1 TeV to 250 TeV are given by
\begin{eqnarray}
 \Phi_{0,\rm{p}}~&=~(4.96 \pm0.53) \tmE{3} \rm{(m^{2}~sr~s)^{-1}~ (GeV/n)^{\alpha_{p}-1},  ~\alpha_{p}~ = 2.61 \pm 0.01~for~protons~and }\\
 \Phi_{0,\rm{He}}~&=~(3.01 \pm0.30) \tmE{3} \rm{(m^{2}~sr~s)^{-1}~ (GeV/n)^{\alpha_{He}-1}, ~\alpha_{He}~ = 2.55 \pm 0.01~for~helium~nuclei.}
\end{eqnarray}

\begin{figure}
\centering
 \includegraphics[width=0.48\textwidth, trim= 0 10 0 10]{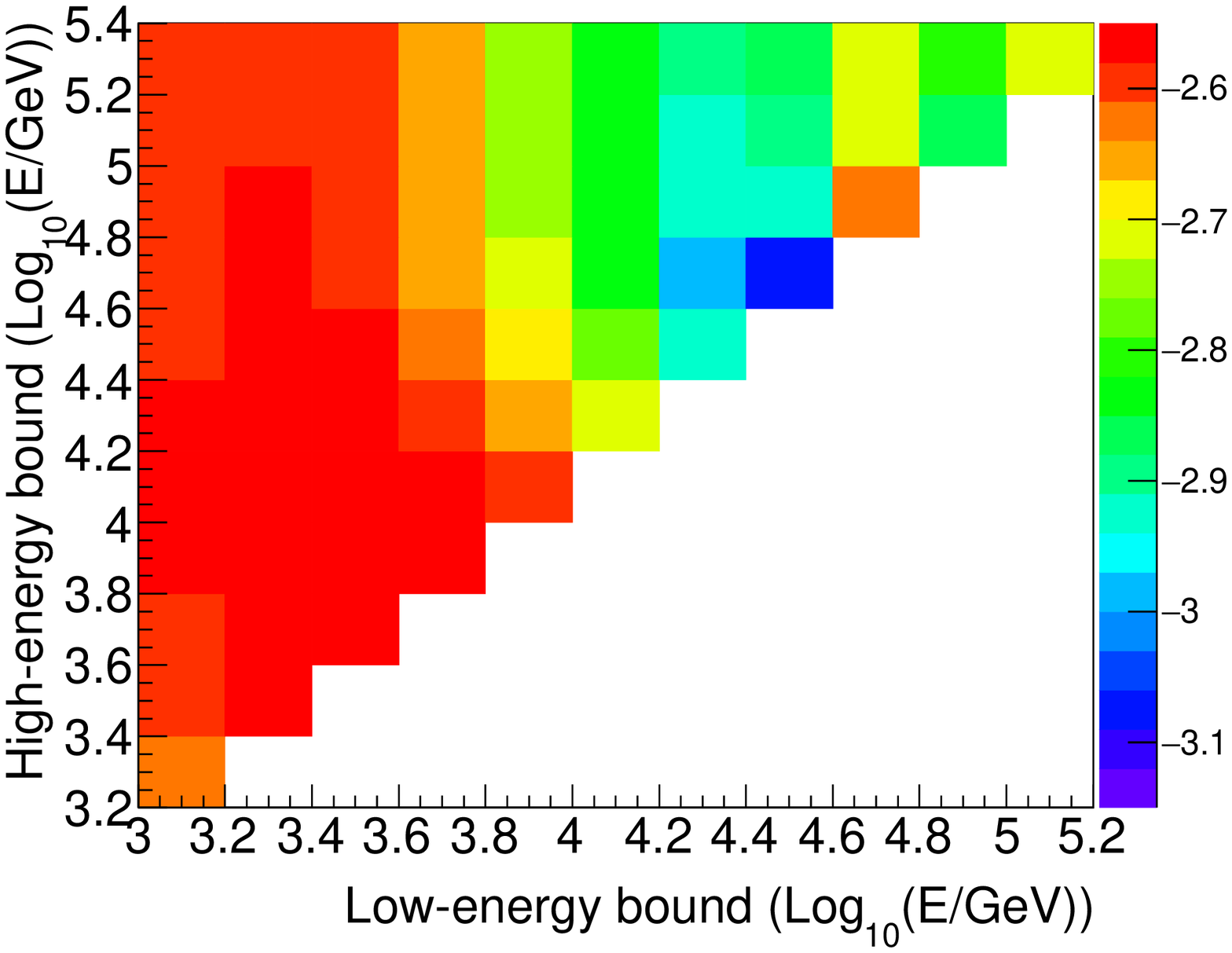}
 \includegraphics[width=0.48\textwidth, trim= 0 10 0 10]{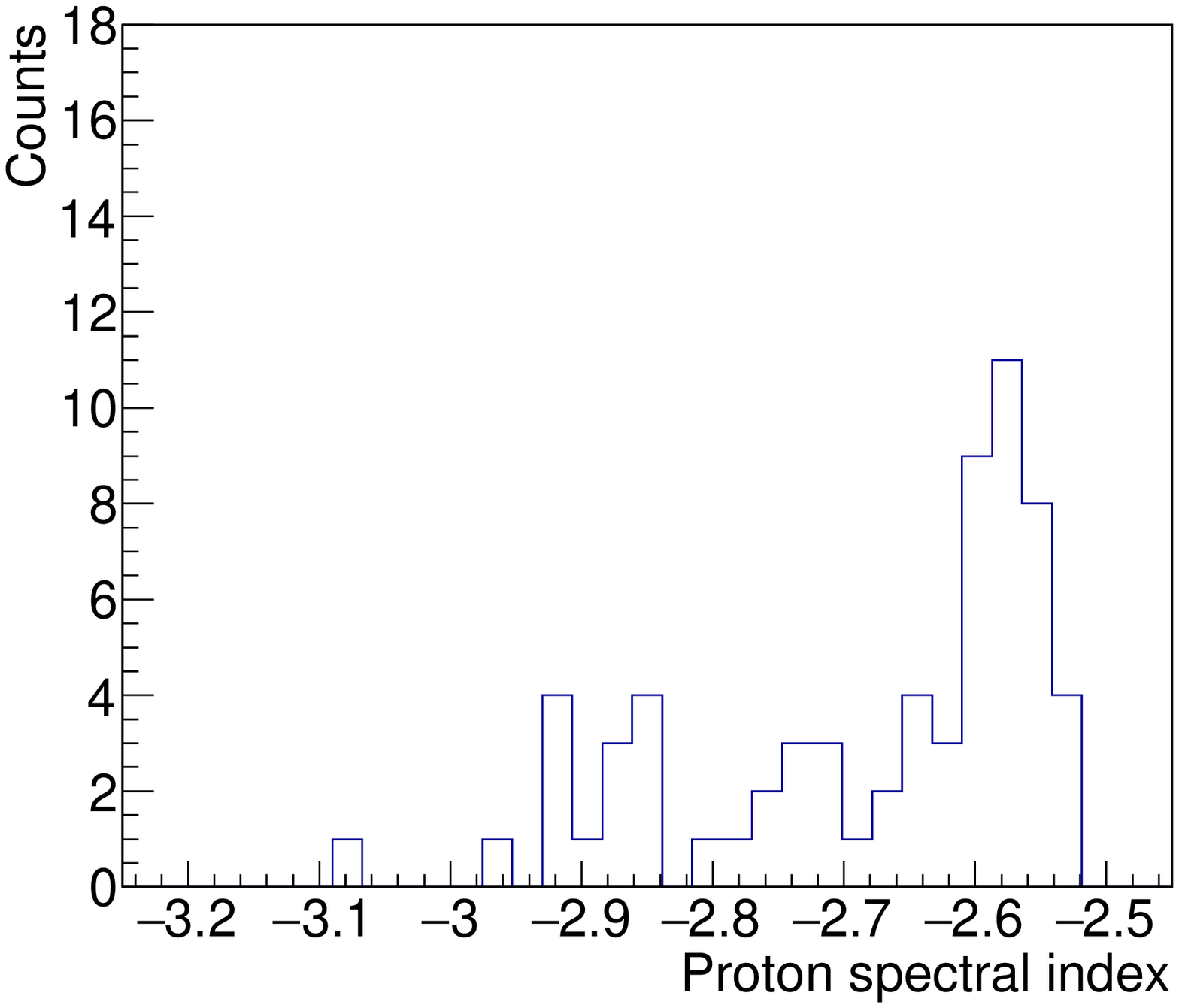}\\
\caption
{Distribution (left) of the best-fit spectral index as a function of the low- and high-energy bounds for the range of points used in the fit, 
for CREAM-III proton data, and distribution (right) of the resulting best-fit index values.}
\label{fig:protonfits}
\end{figure}

\begin{figure}
\centering
 \includegraphics[width=0.48\textwidth, trim= 0 10 0 10]{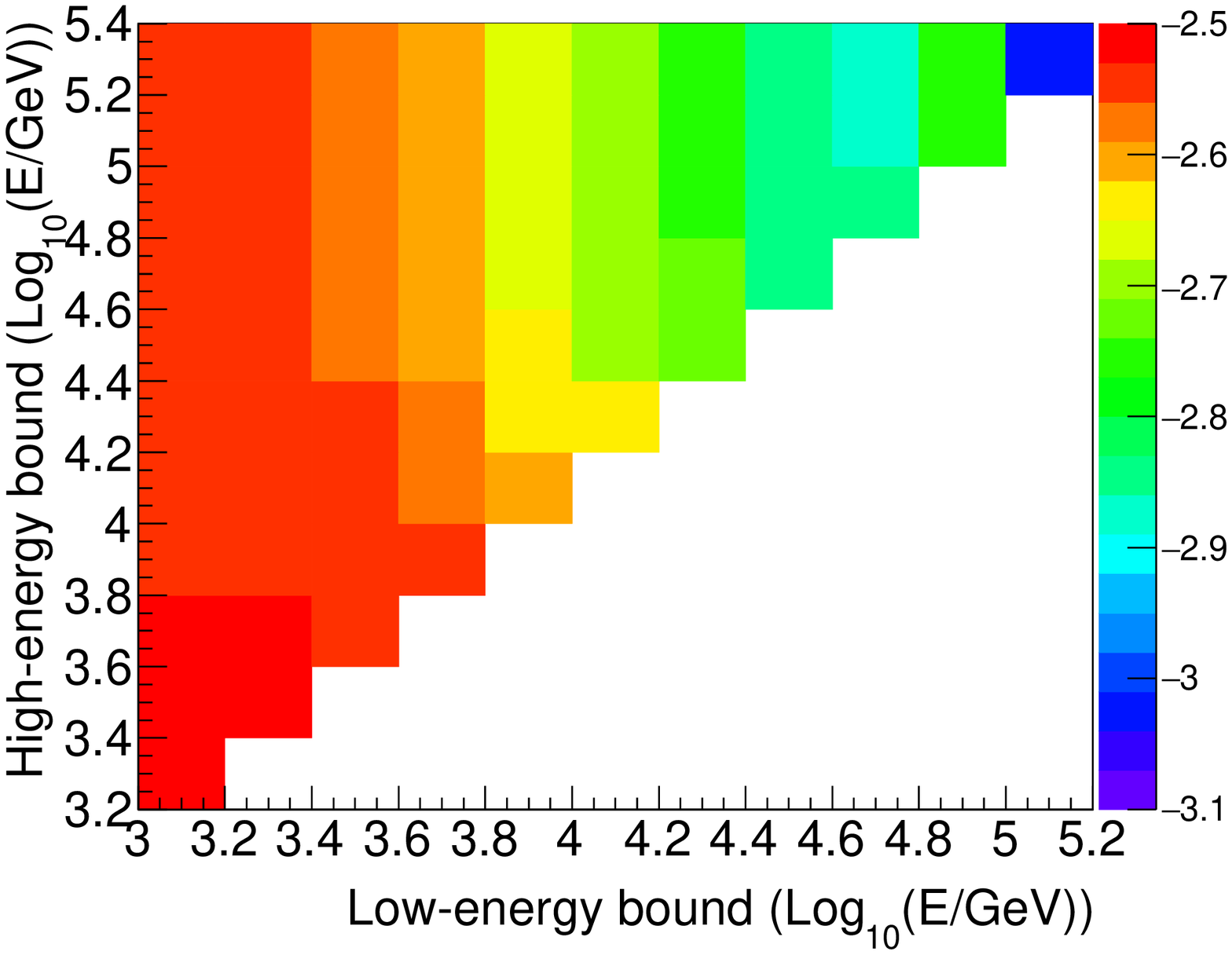}
 \includegraphics[width=0.48\textwidth, trim= 0 10 0 10]{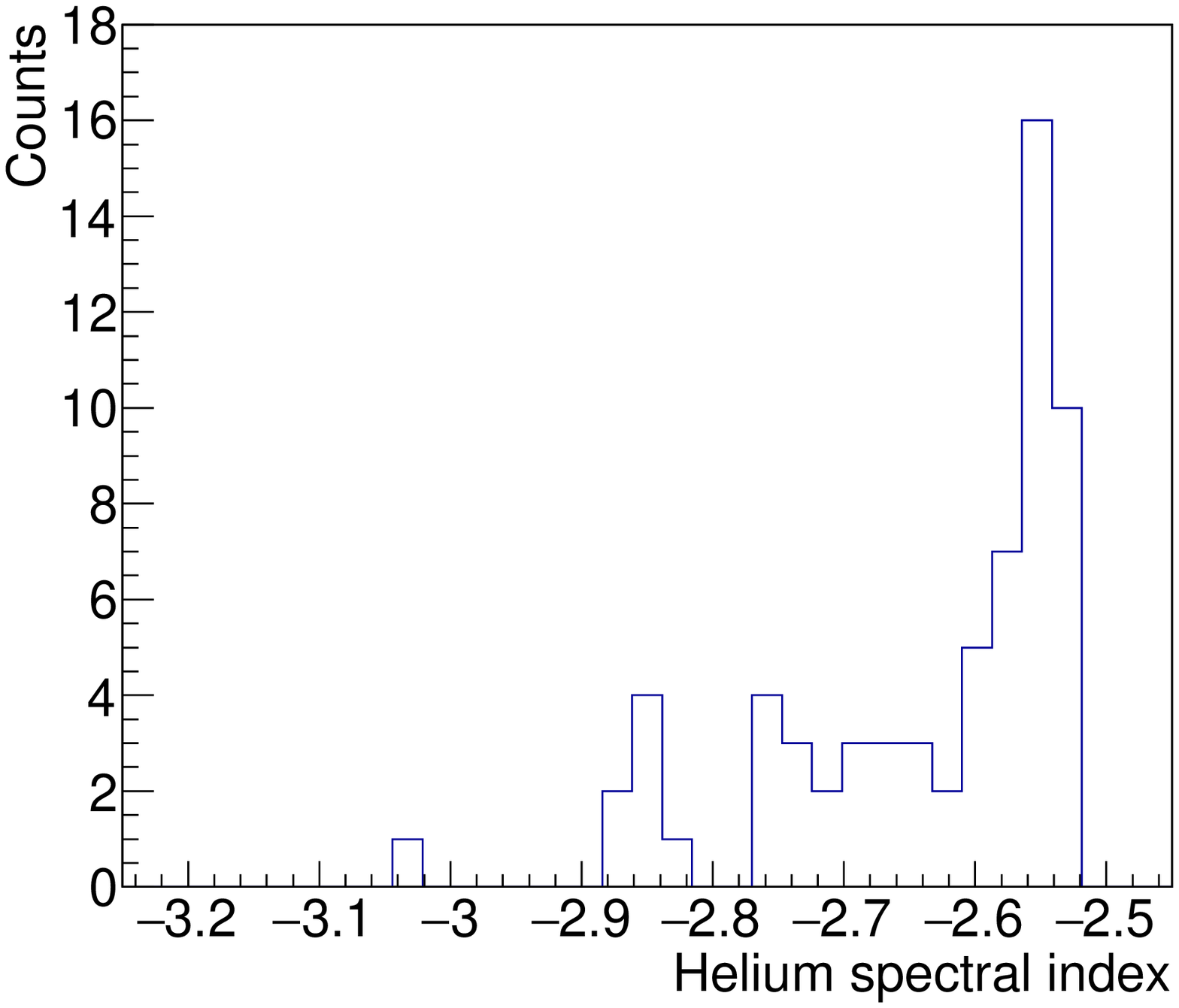}\\
\caption
{Distribution (left) of the best-fit spectral index as a function of the low- and high-energy bounds for the range of points used in the fit, for CREAM-III helium data, and distribution (right) of the resulting best-fit index values.}
\label{fig:heliumfits}
\end{figure}

The indices for the CREAM-III proton and helium spectra over the CREAM-I measurement range, 2.5 TeV to 250 TeV, 
are 2.65 $\pm$ 0.03 and 2.60 $\pm$ 0.03, respectively, 
while the reported indices for the CREAM-I proton and helium spectra are 2.66 $\pm$ 0.02 and 2.58 $\pm$ 0.02, respectively. 
Thus the CREAM-I and CREAM-III results are comparable within uncertainties.
The CREAM-III proton and the helium fluxes above \sm20 TeV appear systematically lower 
than extrapolated fluxes from 1 to 10 TeV, although at the higher energies the fluxes have large statistical uncertainties. 
Fig. \ref{fig:protonfits} and Fig. \ref{fig:heliumfits} show distributions of best-fit spectral indices for proton and helium nuclei, respectively,  
as a function of low- and high-energy bounds used in selecting the data input to the fit.
These figures show a gradual change of index as the low-energy bound is increased, for both proton and helium spectra.
The large similarly-colored area in the 2-dimensional maps correspond 
to a most probable index value of about -2.6 for proton and helium spectra, in the region below 20 TeV. 

\begin{figure}
\centering
 \includegraphics[width=0.48\textwidth, trim= 10 10 110 10]{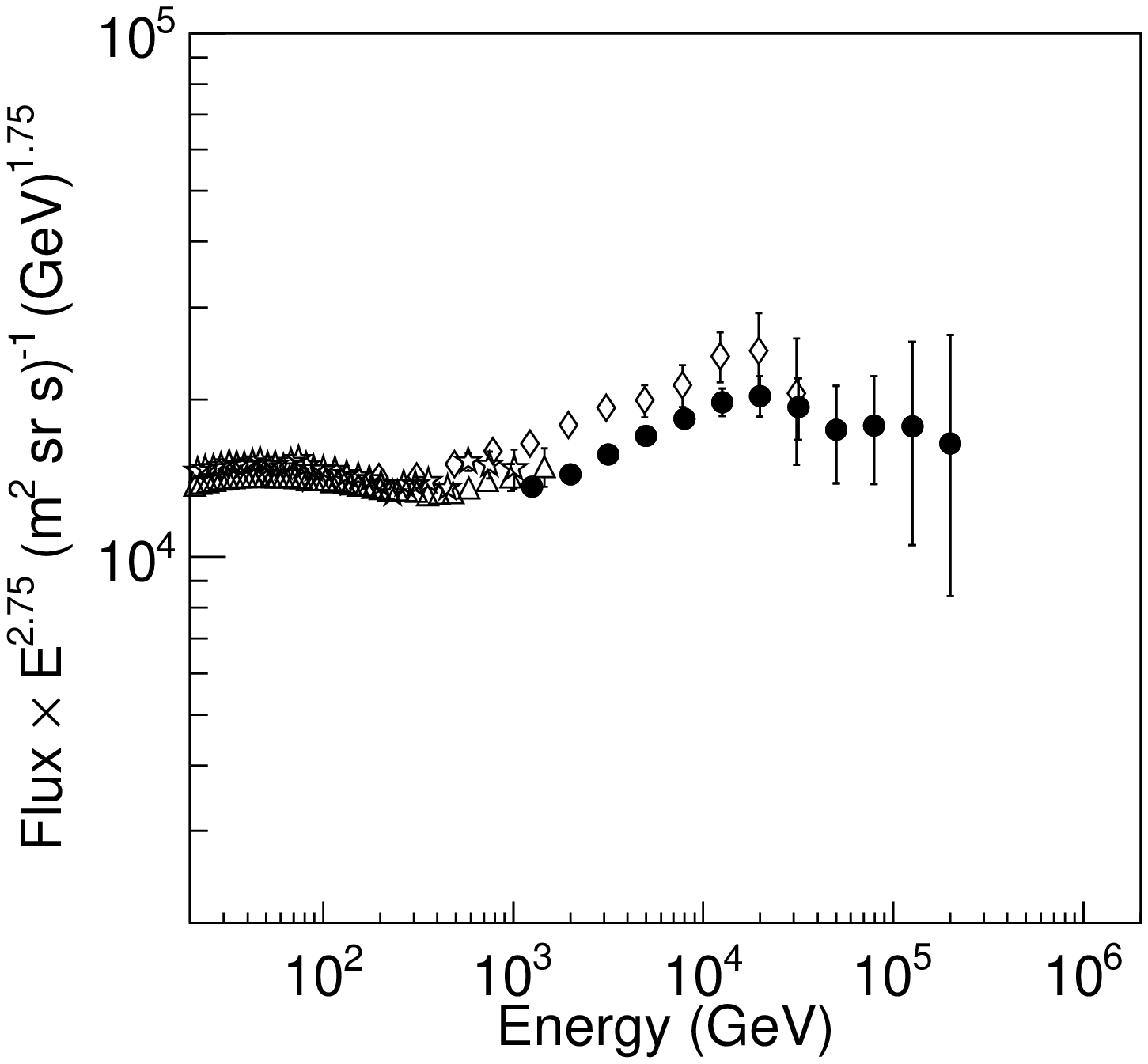}
 \includegraphics[width=0.48\textwidth, trim= 10 10 110 10]{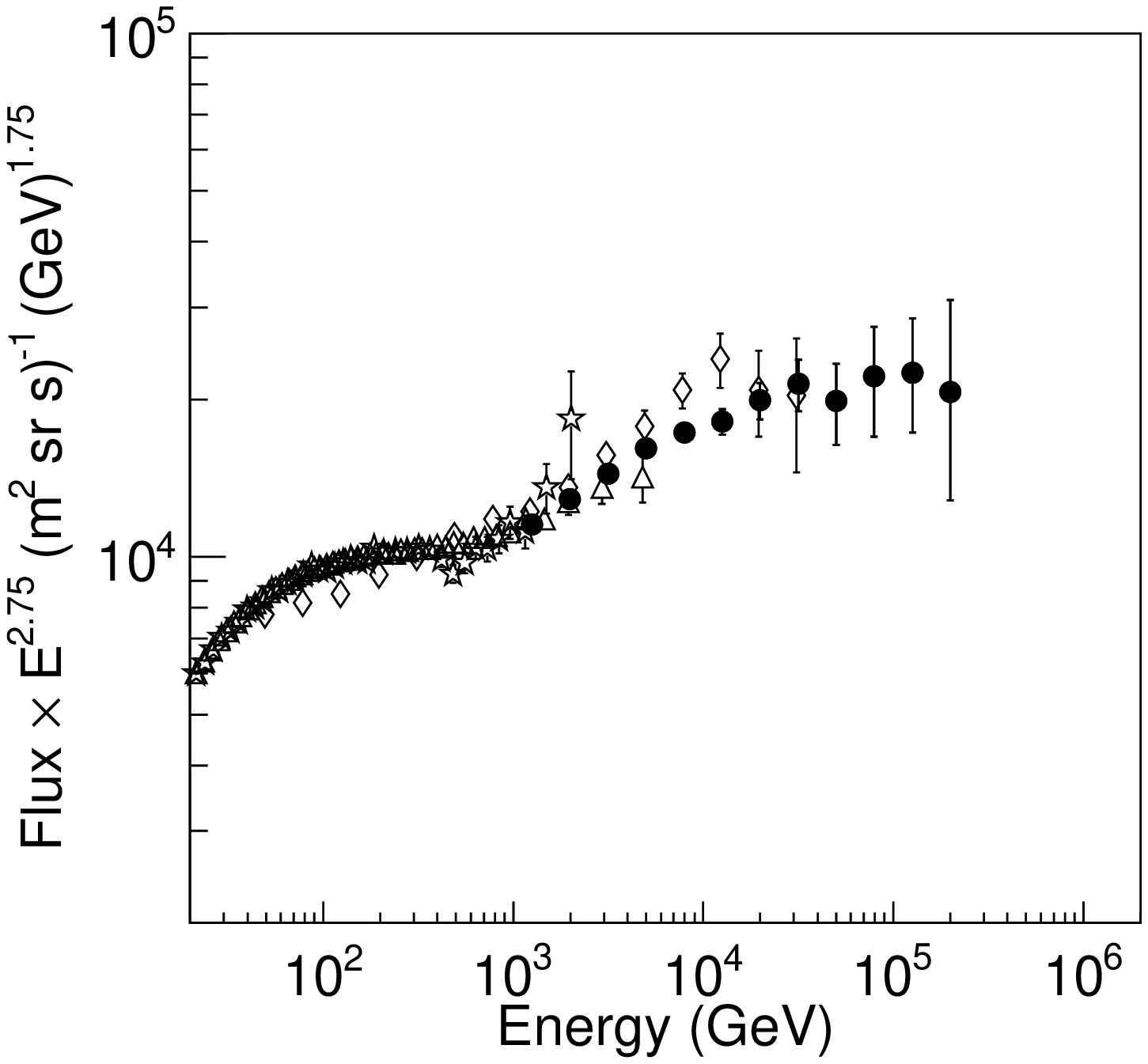}\\
\caption
{Proton (left) and helium (right) spectra from the combined CREAM-I and CREAM-III data (filled circles). 
Statistical uncertainties are shown. 
Selected previous measurements are also shown: AMS-02 (triangles), ATIC-2 (diamonds), and PAMELA (stars).}
\label{fig:hheflux13}
\end{figure}

The combined fluxes from CREAM-I and CREAM-III for protons and helium nuclei, 
given in Table \ref{tab:cream13p} and shown in Fig. \ref{fig:hheflux13} 
were estimated by weighting the separate spectra with their respective flight exposures.
Uncertainties are propagated from the separate statistical uncertainties of both results. 
As shown in Fig. \ref{fig:hheflux13}, the CREAM-I and CREAM-III combined data show the same steepening as the CREAM-III data.
However, their statistical uncertainties are rather large above \sm20 TeV, and more data are needed.

\floattable
\begin{deluxetable}{ccc|ccc}
\tablecaption{Combined CREAM-I and CREAM-III fluxes for protons and helium nuclei \label{tab:cream13p}}
\tablecolumns{9}
\tablenum{3}
\tablewidth{0pt}
\tablehead{\colhead{$E_{i}$} &
\colhead{$E_{f}$} &
\colhead{Proton Flux} &
\colhead{$E_{i}$} & 
\colhead{$E_{f}$} &
\colhead{Helium Flux}\\
\colhead{(GeV) } &
\colhead{ (GeV) } &
\colhead{((m$^{2}$s sr GeV)$^{-1}$)} &
\colhead{(GeV n$^{-1}$) } & 
\colhead{(GeV n$^{-1}$) } &
\colhead{(m$^{2}$ s sr GeV n$^{-1}$)$^{-1}$}  }
\startdata
$10^{3}$    & 1.58\tmE{3} & $ ( 4.07 \pm 0.05) \times 10^{-5} $ & 
2.51\tmE{2} & 3.98\tmE{2}     & $ (1.38 \pm 0.02 ) \times 10^{-4} $ \\
1.58\tmE{3} & 2.51\tmE{3} & $ ( 1.21 \pm 0.02 ) \times 10^{-5} $ &
3.98\tmE{2} & 6.31\tmE{2}     & $ (4.35 \pm 0.08 ) \times 10^{-5} $ \\
2.51\tmE{3} & 3.98\tmE{3} & $ ( 3.72 \pm 0.07  ) \times 10^{-6} $ &
6.31\tmE{2} & $10^{3}$ 	         & $ (1.37 \pm 0.03 ) \times 10^{-5} $ \\
3.98\tmE{3} & 6.31\tmE{3} & $ ( 1.14 \pm 0.03 ) \times 10^{-6} $ &
$10^{3}$    & 1.58\tmE{3}       & $ (4.31 \pm 0.12 ) \times 10^{-6} $ \\
6.31\tmE{3} & $10^{4}$    & $ ( 3.46 \pm 0.02 ) \times 10^{-7} $ &
1.58\tmE{3} & 2.51\tmE{3}     & $ (1.30 \pm 0.05 ) \times 10^{-6} $ \\
$10^{4}$    & 1.58\tmE{4} & $ ( 1.05 \pm 0.06 ) \times 10^{-7} $ &
2.51\tmE{3} & 3.98\tmE{3}     & $ (3.85 \pm 0.22 ) \times 10^{-7} $ \\
1.58\tmE{4} & 2.51\tmE{4} & $ ( 3.04 \pm 0.27  ) \times 10^{-8} $ &
3.98\tmE{3} & 6.31\tmE{3}     & $ (1.19 \pm 0.10 ) \times 10^{-7} $ \\
2.51\tmE{4} & 3.98\tmE{4} & $ ( 8.2 \pm 1.1   ) \times 10^{-9} $ &
6.31\tmE{3} & $10^{4}$ 	         & $ (3.60 \pm 0.41 ) \times 10^{-8} $ \\
3.98\tmE{4} & 6.31\tmE{4} & $ 2.1^{+0.5}_{-0.4} \times 10^{-9} $ &
$10^{4}$    & 1.58\tmE{4}       & $  9.4^{+1.7}_{-1.7} \times 10^{-9} $ \\
6.31\tmE{4} & $10^{5}$    & $ 6.0^{+1.6}_{-1.4} \times 10^{-10} $ &
1.58\tmE{4} & 2.51\tmE{4}     & $  3.0^{+0.7}_{-0.7}\times 10^{-9} $ \\
$10^{5}$    & 1.58\tmE{5} & $ 1.7^{+0.8}_{-0.7} \times 10^{-10} $ &
2.51\tmE{4} & 3.98\tmE{4}     & $  8.5^{+2.3}_{-2.0} \times 10^{-10} $\\
1.58\tmE{5} & 2.51\tmE{5} & $ 4.4^{+2.7}_{-2.2} \times 10^{-11} $ &
3.98\tmE{4} & 6.31\tmE{4}     & $  2.2^{+1.1}_{-0.8} \times 10^{-10} $\\
\enddata
\end{deluxetable}

\begin{figure}
\centering
 \includegraphics[width=0.65\textwidth]{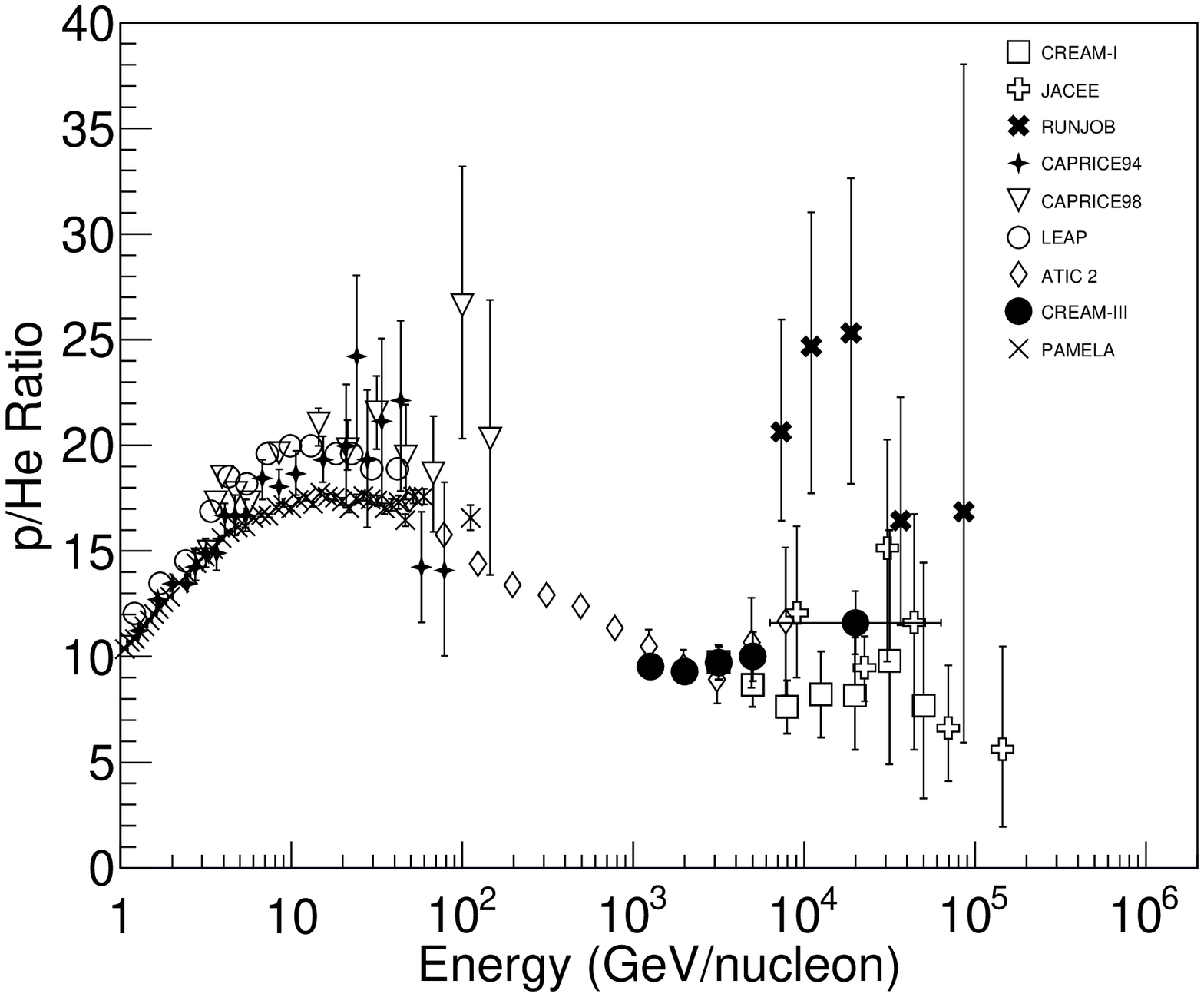}\\
\caption
{Ratio of proton to helium fluxes from CREAM-I (open squares) and CREAM-III (filled circles) compared with other measurements:
JACEE (open crosses, \citealt{1998ApJ...502..278A}), RUNJOB (filled crosses, \citealt{2001APh....16...13A}), 
CAPRICE94 (pluses, \citealt{1999ApJ...518..457B}), CAPRICE98 (open inverted triangles, \citealt{2003APh....19..583B}) , 
LEAP(open circles, \citealt{1991ApJ...378..763S}), ATIC-2 (open diamonds), and PAMELA (thin crosses).
Statistical uncertainties for CREAM-I and CREAM-III data are shown.}
\label{fig:pHeratios}
\end{figure}

The measured CREAM-III p/He ratio is on average 9.6 $\pm$ 0.3 for the 1 \TeVN$~$to 63 \TeVN~range, 
while the CREAM-I ratio was 9.1 $\pm$ 0.5 from 2.5 \TeVN$~$to 63 \TeVN. 
The measured ratio from CREAM-III is significantly smaller than that of \sm20 measured in the range 10-100 GeV/n (See Fig. \ref{fig:pHeratios}).  
The p/He ratios in CREAM-III have large statistical uncertainties and fluctuations above 10 \TeVN,
where more data are clearly needed.
 
\section{Discussion}
\begin{figure}
\centering
 \includegraphics[width=0.65\textwidth]{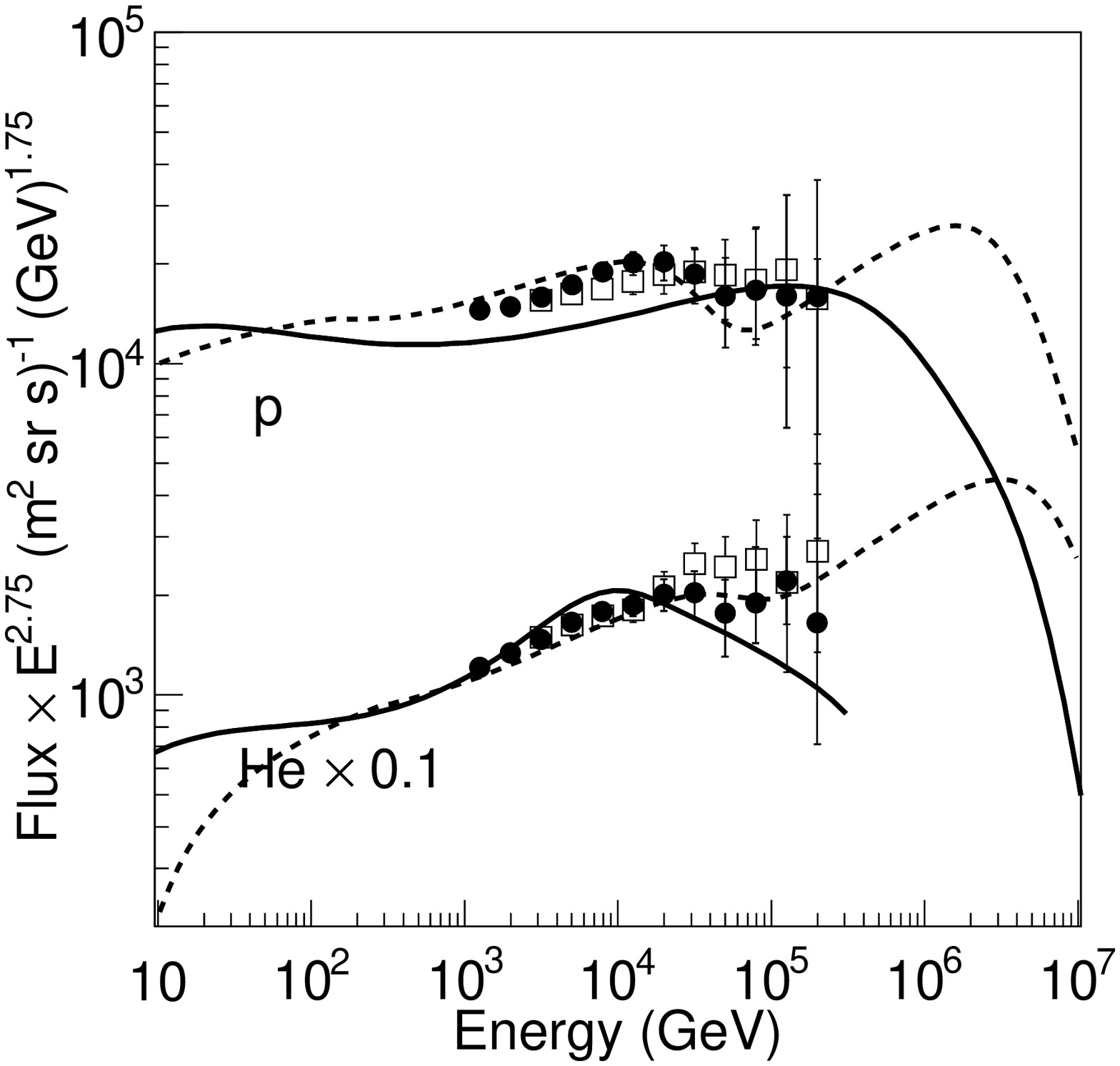}\\
\caption
{CREAM-III (filled circles) and CREAM-I (open squares) spectra for proton and helium nuclei, compared with models by \citeauthor*{2006AA...458....1Z} (2006, dashed line) and \citeauthor*{2010ApJ...718...31P} (2010, solid line)}
\label{fig:models}
\end{figure}

As mentioned in Sec. \ref{sec:intro}, a number of models were suggested to explain the observed hardening and abundances in cosmic-ray spectra.
The CREAM-III results are compared here with two models by \citet{2006AA...458....1Z} and \citet{2010ApJ...718...31P},
as shown in Fig.  \ref{fig:models}.
The model of \citet{2006AA...458....1Z} describes the cosmic-ray spectrum in terms of three different classes of sources,
each indicating a power-law in rigidity with its specific spectral index and maximal rigidity. 
For example, it is assumed that the first class comes from the explosion of isolated stars into the interstellar medium, 
the second one comes from supernovae within the local super-bubble, 
and the third one comes from nova stars with consideration of solar modulation effects characterized by the modulation parameter. 
The first class of sources is limited in its energy reach,
while sources from the second class accelerate cosmic rays efficiently up the knee region at about 3\tmE{15} eV,
which results in a softening over tens of TeV.
The contribution of third class sources is mostly in the energy region below $\sim$ 300 \GeVN.
In this model, the hardening of proton and helium spectra in the TeV regime is accounted for from spectral index differences
between the first and third classes,
i.e., supernovae into the interstellar medium and novae, respectively.
The other model in Fig. \ref{fig:models} by \citet{2010ApJ...718...31P} is a steady-state cosmic-ray spectrum, 
taking into account magnetic-field amplification and Alfvenic drift both upstream and downstream of supernova shocks.  
Different types of supernova remnants (SNRs) and their evolution were considered, 
including type Ia, type IIP, type Ib/c, and type IIb SNRs, which constitute about 90\% of all supernovae. 
The source normalization for nuclei from protons to iron was adjusted from a fit to the observed cosmic-ray composition at a reference energy of 1 TeV. 
According to the modeling, the Type IIP SNRs (most frequent in the Galaxy) accelerate particles by forward shocks up to about 100 TeV and 
the Type Ia and Type Ib/c SNRs (less frequent) accelerate particles up to the knee. 
The reverse shocks in theses SNRs produce very hard particle spectra with about order of magnitude smaller maximum energies.
Each shock is modified by the pressure of the accelerated particles and produces the concave energy spectrum.
This explains the deviations of the calculated interstellar cosmic-ray spectrum from a simple power-law form.

Measured proton and helium energy spectra from the CREAM-III flight exhibit significant fluctuations 
as well as an apparent suppression beyond 20 TeV, where statistical uncertainties are large. 
Additional data will be needed to reduce statistical uncertainties and clarify the situation, 
especially above 100 TeV, where there are few events for previous instrument exposures.
The CREAM payload has recently been transformed for accommodation on the International Space Station (ISS). 
It is currently scheduled to launch on SpaceX-12 in June 2017 to be installed on the ISS JEM-EF module. 
The CREAM instrument on the ISS is expected to have better performance than that for the balloon flights.
For instance, the silicon charge detector (SCD) provides four independent measurements of the charge, each with a resolution of $<$0.2$e$, 
while the SCD on the balloon flights provided two independent measurements. 
In addition, reduced event losses are expected on the ISS compared to the balloon flights,
due to reduced atmospheric overburden and detector materials above the SCD.
The performance of the calorimeter on the ISS is expected be the same as that for the balloon flights, 
measuring cosmic rays in the energy range of 10$^{11}$ to  $>10^{15}$ eV, with energy resolution $\sim$ 40$\%$.
Along with newly introduced top and bottom counting detectors and boronated scintillator detector,
it is expected to measure lower energy events on the ISS, especially electrons given a new e/p separation capability.
The CREAM energy reach is expected to be increased by more than an order of magnitude with a three-year exposure on the ISS.
A 3-year exposure on the ISS will greatly reduce the statistical uncertainties and 
extend CREAM measurements to energies beyond any reach possible with balloon flights, as illustrated in \citet{2014AdSpR..53.1451S}.

\acknowledgments

This work was supported in the U.S. by NASA grants NNX11AC50G, NNX11AC52G, and their predecessor grants, 
in Mexico by DGAPA-UNAM grant IN109617,
and in Korea by the National Research Foundation grants (No. 2015R1A2A1A01006870, and No. 2015R1A2A1A15055344). 
The authors thank NASA Wallops Flight Facility, Columbia Scientific Balloon Facility, and 
National Science Foundation Office of Polar Programs for the successful balloon launch, flight operations, and payload recovery. 
This work is supported by NASA in the U.S., KICOS and Ministry of Science and Technology in Korea, and IN2P3, CNRS, and CNES in France.

\bibliography{CREAMIIIpHe}{}
\bibliographystyle{aasjournal}

\listofchanges

\end{document}